\documentclass[english,prapplied,aps,reprint,superscriptaddress,twocolumn]{revtex4-2}
\usepackage[T1]{fontenc}
\usepackage{units}
\usepackage{bm}
\usepackage{amsmath}
\usepackage{amssymb}
\usepackage{graphicx}
\usepackage{esint}
\usepackage{booktabs}

\makeatletter
\usepackage[english]{babel}
\usepackage{graphicx}
\usepackage{dcolumn}
\usepackage{bm}
\usepackage{xcolor}

\def\url#1{}
\addto\captionsenglish{%
}

\usepackage{babel}

\providecommand*{\diff}%
{\@ifnextchar^{\DIfF}{\DIfF^{}}}
\def\DIfF^#1{%
	\mathop{\mathrm{\mathstrut d}}%
	\nolimits^{#1}\gobblespace}
\def\gobblespace{%
	\futurelet\diffarg\opspace}
\def\opspace{%
	\let\DiffSpace\!%
	\ifx\diffarg(%
	\let\DiffSpace\relax
	\else
	\ifx\diffarg[%
	\let\DiffSpace\relax
	\else
	\ifx\diffarg\{%
	\let\DiffSpace\relax
	\fi\fi\fi\DiffSpace}
\usepackage{babel}

\makeatother

\begin{document}

\title{Fast \textit{ab initio} design of high-entropy magnetic thin films}
\author{Dinesh Bista}
\affiliation{Department of Physics, Georgetown University, Washington, D.C. 20057,
USA}
\author{Willie B. Beeson}
\affiliation{Department of Physics, Georgetown University, Washington, D.C. 20057,
USA}
\author{Turbasu Sengupta}
\affiliation{Department of Physics, Virginia Commonwealth University, Richmond, VA, 23284,
USA}
\author{Jerome Jackson}
\affiliation{Scientific Computing Department, STFC Daresbury Laboratory, Warrington WA4 4AD, United Kingdom}
\author{Shiv N Khanna}
\affiliation{Department of Physics, Virginia Commonwealth University, Richmond, VA, 23284,
USA}
\author{Kai Liu}
\affiliation{Department of Physics, Georgetown University, Washington, D.C. 20057,
USA}
\author{Gen Yin}
\thanks{gen.yin@georgetown.edu}
\affiliation{Department of Physics, Georgetown University, Washington, D.C. 20057,
USA}
\begin{abstract}
We show that the magnetic properties of high-entropy alloys (HEAs) can be captured by \textit{ab initio} calculations within the coherent potential approximation, where the atomic details of the high-entropy mixing are considered as an effective medium that possesses the translational symmetry of the lattice. This is demonstrated using the face-centered cubic (FCC) phase of $\textrm{FeCoNiMnCu}$ and the $L1_0$ phase of $\textrm{(FeCoNiMnCu)Pt}$ by comparing the density functional theory (DFT) results with the experimental values. Working within the first Brillouin zone and the primitive unit cell, we show that DFT can capture the smooth profile of magnetic properties such as the saturation magnetization, the Curie temperature and the magnetic anisotropy, using only a sparse set of sampling points in the vast compositional space. The smooth profiles given by DFT indeed follow the experimental trend, demonstrating the promising potential of using machine learning to explore the magnetic properties of HEAs, by establishing reasonably large datasets with high-throughput calculations using density-functional theory. 

\end{abstract}
\maketitle
 High-entropy alloys (HEA) are a category of solids with a long-range ordered crystal structure, whereas the atomic sites are randomly occupied by the homogeneous, random mixing of five or more different elements\cite{zhang_microstructures_2014,yeh_formation_2004,chen_nanostructured_2004,cantor_microstructural_2004,yeh_nanostructured_2004,miracle_critical_2017,george_high_2020,george_high-entropy_2019,kozak_single-phase_2015,praveen_high-entropy_2018}.
 After this mixing, the change in the Gibbs free energy $\Delta G=\Delta H-T\Delta S$ is allowed to be negative even with a significant increase in the mixing enthalpy $\Delta H>0$. 
 This is enabled by the large configurational entropy gain $\Delta S>0$.
 Such mixing can thus spontaneously stabilize a crystal structure that hosts many chemical bonds normally not favored in regular intermetallic alloys or ordered crystals\cite{yeh_formation_2004}. 
 Since the mixing is homogeneous and random, the stabilized crystal is often described as a solid solution\cite{yeh_nanostructured_2004}. 
 Unlike regular ordered alloys, solid solutions host random local strains and lattice distortions because of the complex, multi-element composition. 
 Therefore, atomic diffusion, dislocation, and grain boundary movements are impeded by numerous obstacles. 
 This leads to increased hardness, thermal stability, and reduced brittleness, while surprisingly maintaining good ductility\cite{lei_enhanced_2018,li_mechanical_2021}. 
 Similar to the intriguing mechanical properties, solid-solution phases of HEA are also known to possess high chemical stability\cite{chen_microstructure_2005,lee_effect_2007,george_high-entropy_2019}. 
 Due to the uniform mixing of the high-entropy atomic species, phase segregation in HEAs is more suppressed compared to regular alloys. 
 This impedes the progress of electrochemical processes and thereby enhances the robustness against bimetallic corrosion. 
 Due to these intriguing properties, HEAs have undergone extensive investigation, primarily focused on their enhanced mechanical and chemical properties \cite{law_review_2023,vaidya_high-entropy_2019,ye_high-entropy_2016,miracle_critical_2017}. 
 Recent developments in machine learning and autonomous workflow have brought both the experimental and theoretical research to a new level\cite{rao_machine_2022,divilov_disordered_2024,huang_machine-learning_2019,oses_aflow-chull_2018-1}. 
In addition to mechanical and chemical robustness, HEAs can also exhibit exceptional magnetic properties\cite{wang_microstructure_2014,liu_microstructure_2012,law_review_2023,tang_magnetism_2021}. 
The ordered structures of HEAs naturally break the $SO(3)$ symmetry, allowing for good magnetic anisotropy. 
Also, the random local distortion of the lattice can serve as pinning sites, resulting in good permanent magnets with large coercivity\cite{na_hard_2021}. 
Due to the compositional complexity, it is possible to modulate the key magnetic properties such as the saturation magnetization ($M_S$), the magnetic anisotropy energy density (MAE) and the Curie temperature ($T_C$) in a reasonably large range\cite{kumari_comprehensive_2022,ye_high-entropy_2016}. 
This provides the opportunity to identify ideal media for advanced technologies such as heat- or microwave-assisted magnetic recording (HAMR and MAMR), where fine-tuning of $M_S$, MAE and $T_C$ are needed to identify the best balance among the recording trilemma for each generation of the technology\cite{weller_thermal_1999,rottmayer_heat-assisted_2006,weller_review_2016,davies_magnetization_2004,rahman_controlling_2009,gilbert_tuning_2013,cuadrado_-planeout--plane_2016,weller_review_2016}. 
This calls for the investigation of magnetic properties of thin-film HEAs in their vast, high-dimensional compositional space. 
Here, we demonstrate the design principle for magnetic properties of thin-film HEAs using density-functional theory (DFT). 
By comparing to experimental results, we demonstrate that bulk DFT calculations using thin-film lattice constants can capture the smooth profile of magnetic properties in the high-entropy compositional space. 
Specifically, we demonstrate this design principle for the MAE of two thin-film HEAs that have been recently achieved experimentally: the face-centered cubic (FCC) phase of $\textrm{FeCoNiMnCu}$ and the $L1_0$ phase of $\textrm{(FeCoNiMnCu)Pt}$, where robust solid solutions with large MAE and exceptional long-range crystallization have been identified\cite{beeson_single_2023}. 
Particularly, we show that the first-principles calculation can be carried out efficiently at the level of coherent-potential approximation (CPA), a Green's function-based theory for the electronic structure of a lattice featuring compositional and magnetic disorder, allowing for fast and accurate evaluation of magnetic properties of random alloys without the necessity of sampling different configurations using supercells\cite{soven_coherent-potential_1967,soven_contribution_1969,Woodgate_Cantor-Wu_2023}. 
This illustrates the possibility of using machine-learning models to make reasonable predictions once a large-scale database is established. 
\begin{figure*}
\begin{centering}
\includegraphics[width=1\textwidth]{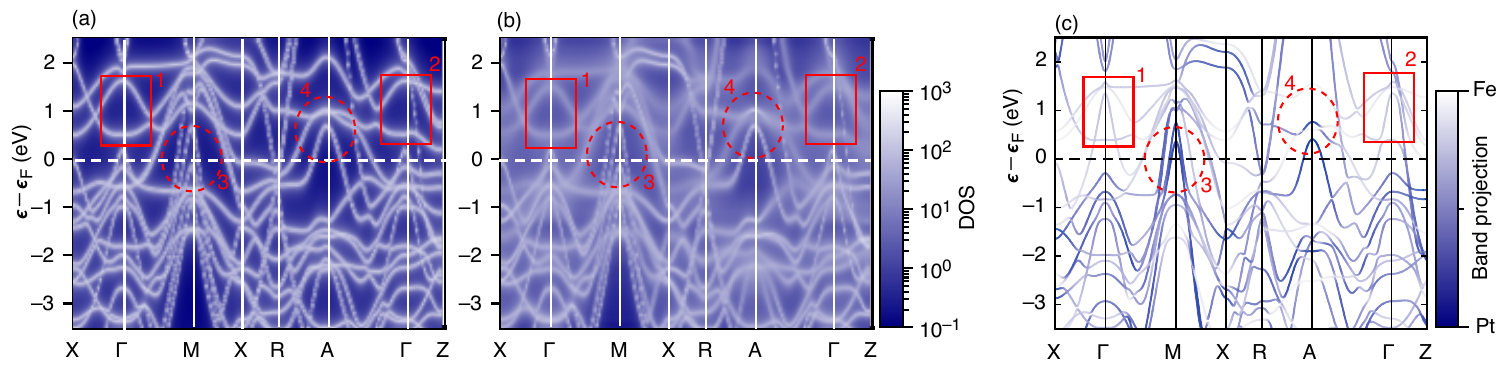}
\end{centering}
\centering{}\caption{(a) The Spectral function of $L1_0\thinspace\textrm{-phase}$ $\textrm{FePt}$ obtained by the Green's function implementation of DFT in Questaal. Coherent potential approximation was used assuming no random mixing to resolve the band structure. An overall broadening energy of $0.001\thinspace\textrm{Ryd}$ was applied to visualize the spectral function. (b) The spectral function in the case of $[\textrm{Fe}_{0.8}\textrm{CoNiMnCu}_{0.2}]\textrm{Pt}$ with the random mixing slightly turned on for Fe sites. (c) The VASP result of pure FePt with eigenstates projected to Fe and Pt orbitals. The weight of the projection is denoted by the color scale. The energy cutoff was set to $400\thinspace\textrm{eV}$, and the DFT-D3 dispersion correction with Becke-Johnson damping was used\cite{grimme_a_consistent_2010,becke_a_density-functional_2005,grimme_effect_2011}. A $\Gamma$-centered $12\times12\times9$ mesh was used for the k-space integration. The convergence criterion was set to $10^{-7}\thinspace\textrm{eV}. $\label{fig:Spectral-function}}
\end{figure*}
The \textit{ab initio} calculations in this work are performed using the Green's function formulation of DFT and Linear Muffin Tin Orbitals within the atomic-sphere approximation (ASA). 
To model the irregularities in HEAs, we perform spin-polarized self-consistent calculations using the CPA implemented by the `lmgf' package in the Questaal suite \cite{pashov_questaal_2020}. 
This approximation considers the irregularities as an averaged effective medium with the same translational symmetry of the lattice, allowing us to work within the first Brillouin zone.
%
%
The electron density is obtained by integrating over a $\Gamma$-centered $16\times16\times16$ k-mesh\cite{monkhorst_special_1976} and $25$ energy points uniformly distributed along an elliptical contour of $0.4$ eccentricity, covering the energy range of $0.85\thinspace\textrm{Ryd}$ below the Fermi level $\epsilon_F$. 
The value of $\epsilon_F$ is determined by the charge-neutrality condition as implemented in the package. 
We determine the ground-state magnetic order by exploring all possible spin configurations, treating positive and negative spins along the quantization axis as two independent CPA species. 
%
%
The third-order
potential function approximation ($p_{3}$) and the convergence criterion of $10^{-6}\thinspace\textrm{Ryd}$ are used.
%
%
For the calculations
of saturation magnetization, the magnetic moment per unit cell
in the converged ground state is taken, and then converted to the units of $\textrm{emu/cm}^3$ 
or $\textrm{emu/g}$. 
Spin-orbit coupling is considered as a $\langle L\cdot S\rangle$ term added to the one-body Hamiltonian when calculating the MAE. 
The values of MAE are extracted by comparing the converged ground-state energy and another single-step energy using the converged moments in the previous step, but rotating the spin quantization axis from the easy axis to a perpendicular direction. 
The specific choices of quantization directions are made according to the setup of experiments. 
%
%

We first calibrate our calculations by aligning Questaal results with other DFT implementations and experimental data.  
This comparison focuses on the well-known high anisotropy single crystal $\textrm{FePt}$ in the $L1_0$ phase.
The Questaal calculation was carried out within CPA, assuming the $100\%$ occupation on both the Fe and Pt sites. 
The primitive cell of the $L1_0$ phase is tetragonal, with Fe atoms occupying the corners and the Pt atom at the body center. 
We used $a=2.69\thinspace\textrm{\AA}$ and $c=3.69\thinspace\textrm{\AA}$ as the lattice constants. 
Perdew-Burke-Ernzerhof (PBE) type of generalized gradient approximation (GGA) was used as the exchange-correlation functional\cite{perdew_generalized_1996}. 
The spectral function $A(\epsilon, \mathbf{k})$ along the high-symmetry route is illustrated in Fig. \ref{fig:Spectral-function}(a).  
This electron spectrum changes significantly when the high-entropy random mixing was slightly turned on for the Fe sites, that is, reducing the probability of the Fe occupation to $0.8$ while simultaneously turning on the probability of $(\textrm{CoNiMnCu})_{0.2}$ accordingly.  
To fully understand the band broadening, we further performed a similar calculation using projector augmented wave pseudopotential method\cite{blochl_projector_1994,kresse_from_1999} implemented by Vienna Ab initio Simulation Package (VASP)\cite{kresse_efficiency_1996,kresse_efficient_1996}. 
%
%
%
%
%
The eigenvalues of the Kohn-Sham Hamiltonian are illustrated along the same high symmetry route, as shown in Fig. \ref{fig:Spectral-function}(c). 
%
%
Although there are discrepancies in some details, the Questaal spectrum agrees with the VASP bands generally. 
Due to the random scattering, although all bands are smeared due to the finite life time, different eigenstates have different levels of broadening. 
To show this, four representative spectra are highlighted in Figs. \ref{fig:Spectral-function}(a-c). 
Regions 1 and 2 (red solid boxes) contain Fe-heavy bands, as denoted by the light color in Fig. \ref{fig:Spectral-function}(c). 
As expected, these bands are broadened significantly after turning on the random mixing on the Fe-occupied sites. 
Unlike Regions 1 and 2, Regions 3 and 4 contain several bands that are heavily mixed with the Pt orbitals. 
Specifically, in Region 3, all bands are almost purely given by Pt orbitals, whereas in Region 4, the mixing from Pt orbitals becomes heavy when crossing the A point. 
In these two regions (red dashed ovals), the band smearing mainly occurs for the Fe-dominated bands, whereas the Pt bands remain sharp. 
This is consistent with the fact that the body-center sites are uniformly occupied by Pt atoms without any random mixing. 
After calibrating the band structure, we further calculate the MAE for $L1_0$ FePt and compare the results with experiments, as shown in Table \ref{tab:Comparison}. 
We used both PBE-GGA and local density approximation (LDA)\cite{barth_a-local_1972}.
The calculated $M_S$ from the Questaal matches closely with the experimental
and the VASP results\cite{wolloch_influence_2017}. 
In the case of MAE, it appears that DFT overestimates the MAE by roughly an order of magnitude, which is captured by both VASP and Questaal. 
%

\begin{table}
\centering
\caption{Comparison of MAE and $M_S$ for pure $L1_{0}\textrm{-FePt}$. \label{tab:Comparison}}
\begin{tabular}{ccc} 
\hline 
Method & MAE ($\textrm{ergs/cm}^{3}$) & $M_S$ ($\textrm{emu/cm}^{3}$) \\ 
\hline 
\hline 
Experiment\cite{klemmer_magnetic_1995,weller_high_2000} & $(6.6\sim10)\times10^{7}$ & $1100\pm100$ \\ 
VASP (LDA)\cite{wolloch_influence_2017} & $1.78\times10^{8}$ & $1012$ \\
Questaal (LDA) & $1.65\times10^{8}$ & $1074$ \\
VASP (GGA)\cite{wolloch_influence_2017} & $1.57\times10^{8}$ & $1066$ \\
Questaal (GGA) & $1.56\times10^{8}$ & $1097$ \\
\hline 
\end{tabular}
\end{table}

\begin{figure}
\centering{}\includegraphics[width=1.0\columnwidth]{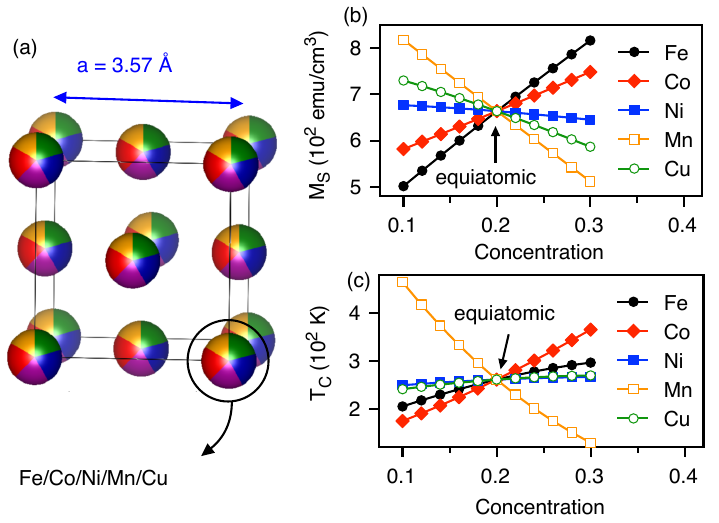} \caption{Magnetic properties of the high-entropy crystal $\textrm{FeCoNiMnCu}$ in the FCC phase. (a) The conventional FCC unit cell with all atomic sites randomly occupied by the five different elements. (b) The saturation magnetization when modulating the concentration of one high-entropy element. All other high-entropy elements are adjusted proportionately to maintain a total concentration of $100\%$. The central composition of $0.2$ corresponds to the equiatomic case. (c) The modulation of Curie temperature corresponding to changes in the concentration of one element, according to the pattern described in (b). }
\label{fig:FCC}
\end{figure}

Following this verification study, we explore the magnetic properties of high-entropy alloys by fully turning on the random mixing. 
We first examine the equiatomic $\textrm{FeCoNiMnCu}$ in the FCC phase, and the results are summarized in Fig. \ref{fig:FCC}.  
The lattice constant of a conventional cell is set to the experimental value $a=3.57\thinspace\textrm{\AA}$, which was previously captured by X-ray diffraction\cite{beeson_single_2023}. 
%
Taking different types of spins for each high-entropy element as separate CPA species, we first examine the ground-state spin configuration of the equiatomic case without considering spin-orbit coupling (SOC). 
After the self-consistent convergence, positive and negative spins of Mn atoms coexist in the ground state, whereas all other elements are ferromagnetically aligned. 
This is consistent with existing theoretical results\cite{rao_unveiling_2020}. 
For the equiatomic composition, LDA gives $M_S=663\thinspace\textrm{emu/cm}^3$, whereas the GGA value is $705\thinspace\textrm{emu/cm}^3$. 
After considering the mass density using the lattice constant, these two values correspond to $81.9\thinspace\textrm{emu/g}$ and $87.1\thinspace\textrm{emu/g}$, respectively. 
These results are close to the DFT (GGA) result ($80.8\thinspace\textrm{emu/g}$) using a similar approach\cite{rao_unveiling_2020}. 


%
To capture the landscape of $M_S$ near the equiatomic case, the compositional space was explored by varying the concentration of one high-entropy element, whereas simultaneously changing the concentration of others accordingly, keeping the total occupation probability normalized. 
The LDA results are shown in Fig. \ref{fig:FCC}(b). 
%
Here, $M_S$ increases with the Fe and Co concentration, whereas extra mixing of Mn and Cu decreases the $M_S$ significantly. 
This is expected since Mn is found to have a mixture of both positive and negative spins, whereas Cu atoms are non-magnetic. 
%
 


\begin{figure}
\centering{}\includegraphics[width=1.0\columnwidth]{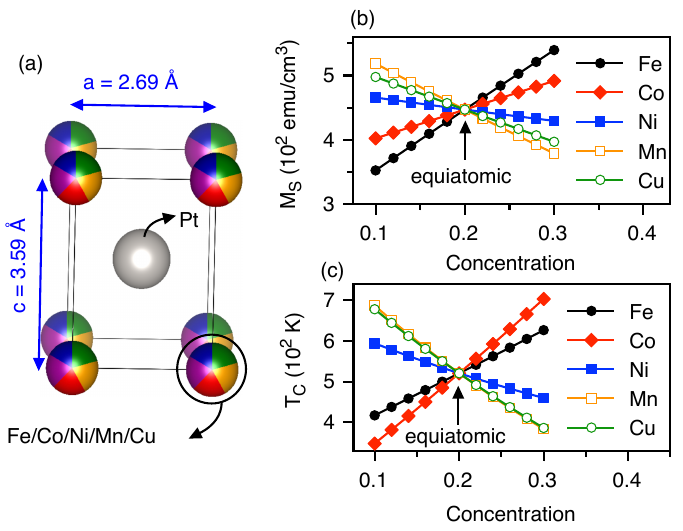} \caption{The saturation magnetization and the Curie temperature of high-entropy magnet  $\textrm{(FeCoNiMnCu)Pt}$ in the $L1_0$ phase. (a) The primitive unit cell of the tetragonal lattice. (b) The saturation magnetization when modulating the concentration of one high-entropy element on the corner sites. All other corner sites are adjusted proportionately to maintain a total concentration of $100\%$. The body-center Pt site is assumed to be uniformly occupied. The $20\%$ concentration corresponds to the equiatomic case for all the corner sites. (c) The modulation of Curie temperature corresponding to changes in the concentration of one element, following the pattern described in (b).  
\label{fig:L10}}
\end{figure}

After exploring the magnetic ordering of equiatomic recipe, we estimate the ordering temperature ($T_C$) through mean-field approximation.\cite{sato_first-principles_2010,kormann_treasure_2015,sato_curie_2003}:
%
%
%
\begin{equation}
T_C=\frac{2(\epsilon_\textrm{PM}-\epsilon_\textrm{GS})}{3(1-c)k_\textrm{B}},\label{eq:Tc}
\end{equation}
where $\epsilon_\textrm{PM}$ and $\epsilon_\textrm{GS}$ represent converged energies of paramagnetic and ground-state spin configurations, respectively. 
Here, $k_\textrm{B}$ is the Boltzmann constant and $c$ denotes the net concentration of the non-magnetic atoms in the HEA.
Note that Eq. \ref{eq:Tc} is a crude approximation for estimating the upper limit\cite{sato_first-principles_2010}. 
%
%
The estimated $T_{C}$ for the equiatomic case using
LDA and GGA are found to be $261\thinspace\textrm{K}$
and $359\thinspace\textrm{K}$, respectively, which are close to existing DFT results\cite{rao_unveiling_2020}. 
We then explore the landscape of $T_C$ in the compositional space near the equiatomic recipe by varying the concentrations similar to that used for $M_S$. 
The LDA results are summarized in Fig. \ref{fig:FCC}(c), where $T_C$ decreases significantly with the increase in the Mn concentration. 
This is consistent with the fact that the ground-state Mn atoms have a mixing of both positive and negative spins, suggesting that their exchange couplings with other high-entropy species should be antiferromagnetic. 
%
%
%

%
To understand the MAE of $\textrm{FeCoNiMnCu}$ in its FCC solid solution phase, we compare the ground-state energy when the spin quantization axis is along different directions. 
Care must be taken since the MAE is only a small fraction of the total energy. 
To evaluate the limit of our numerical resolution, we first examine the energy difference between two equivalent directions within the $\langle100\rangle$ family. 
With a convergence criterion of $10^{-6}\thinspace\textrm{Ryd}$, the energy difference between $[001]$ and $[100]$ corresponds to an MAE of $1.82\times10^{6}\thinspace\textrm{ergs/cm}^3$. 
Although this energy scale is sizable and can indeed be detected experimentally\cite{beeson_single_2023}, it is difficult for DFT calculations to resolve energies that are smaller or near this range. 
We further examine the MAE value when taking $[111]$ and $[110]$ as the easy and hard axes, respectively, which resulted in an MAE of $2.01\times10^{6}\thinspace\textrm{ergs/cm}^3$. 
This is close to the resolution of our calculation, and we therefore conclude that the MAE between $[111]$ and $[110]$ is beyond the numerical accuracy. 
Such a small MAE is a consequence of the highly symmetric cubical lattice, as well as the lack of heavy elements providing a sizable SOC. 
To increase the MAE, we further mix Pt into the FCC solid solution by replacing Mn\cite{kurniawan_curie_2016}, which is known to couple to other high-entropy species antiferromagnetically.  
Keeping the lattice constant the same to the case of $\textrm{FeCoNiCuMn}$, the MAE with the Pt mixing remains close to our numerical resolution ($\sim10^6\textrm{ergs/cm}^3$). 
This suggests that further breaking the symmetries of the cubical lattice may be necessary to increase the MAE. 
One way to break the symmetry of the FCC lattice is to insert a layer of Pt atoms between the high-entropy layers stacking along the $c$ axis. 
The change in the $c$ axis brings the FCC lattice to a tetragonal one, as shown by Fig. \ref{fig:L10}(a).
Pure crystals of this phase is well known to host large MAE in many different cases such as $\textrm{FePt}$, $\textrm{CoPt}$ and $\textrm{MnAl}$\cite{weller_magnetic_1992,sakuma_electronic_1994, lyubina_magnetocrystalline_2005}. 
Recently, the $L1_0$ phase of $\textrm{(FeCoNiMnCu)Pt}$ has been experimentally realized through a co-sputtering followed by a rapid thermal annealing, where both good crystallization and large MAE have been confirmed\cite{gilbert_tuning_2013,gilbert_probing_2014,beeson_single_2023}. 
According to the experimental characterization, we assume that there is no random mixing for the Pt atoms at the body-center sites, whereas all corner sites are occupied by the high-entropy mixing of the transition metal elements. 
This assures that the strong SOC given by the Pt atoms is homogeneously experienced by all the transition metal elements. 
Using the experimental lattice constants listed in Fig. \ref{fig:L10}(a), we explore the magnetic properties of $\textrm{(FeCoNiMnCu)Pt}$ in the compositional space, and the results are shown in Figs. \ref{fig:L10}(b-c).  
Compared to the FCC case, the range of $M_S$ becomes smaller near the equiatomic composition due to the insertion of the non-magnetic $\textrm{Pt}$.  
Unlike the equiatomic FCC case, all compositions we explored for the $L1_0$ phase have a ferromagnetic ground state, resulting in a significant increase of $\sim200\thinspace\textrm{K}$ in $T_C$. 
Increasing the concentrations of Fe and Co increases both $M_S$ and $T_C$, whereas the mixing of Mn and Cu has an opposite trend.
This suggests a potential engineering strategy to reduce the writing power for HAMR devices.  

\begin{figure*}
\centering{}\includegraphics[width=1.0\textwidth]{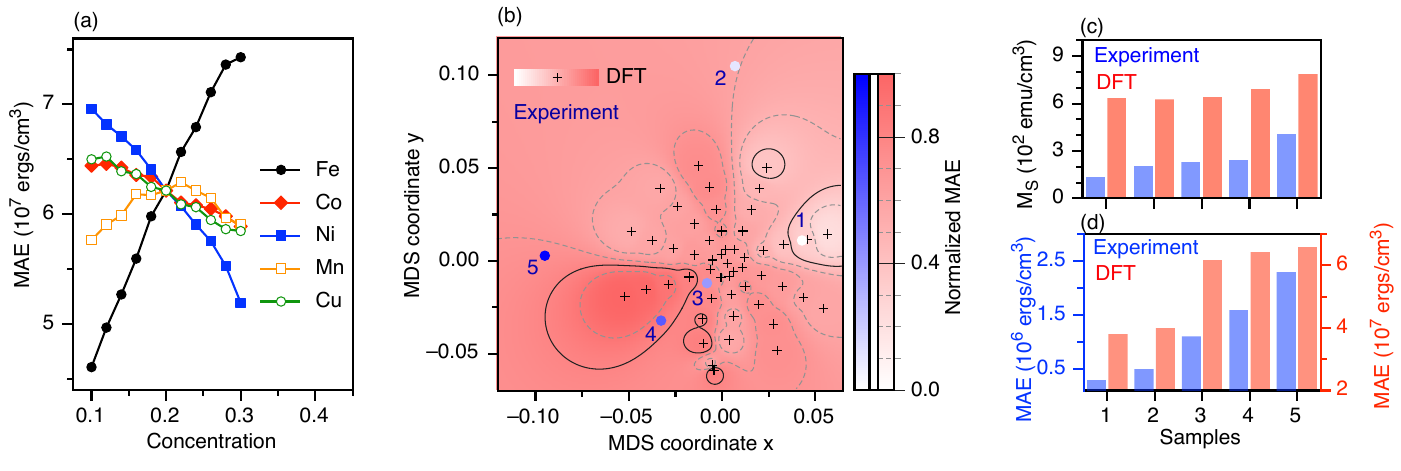} \caption{The magnetic anisotropy energy density (MAE) of the $L1_0$ high-entropy magnet $\textrm{(FeCoNiMnCu)Pt}$. (a) The variation of MAE for different compositions when each high entropy element is modulated following the pattern described in Figs. \ref{fig:FCC} and \ref{fig:L10}.  (b) The compositions in (a) projected from $\mathbb{R}^{5}$ to $\mathbb{R}^{2}$ using Multidimensional Scaling (MDS), which preserves the Euclidean distances between points. Color saturation denotes the MAE values scaled between $0$ and $1$. The red and blue points denote the DFT and the experimental results, respectively. (c-d) Direct comparison between the theoretical and experimental results for the $5$ samples highlighted in (b). Here the $M_S$ comparison is shown in (c), whereas the MAE comparison is shown in (d). Note that $M_S$ is compared with the same scale in (c), whereas the MAE comparison is illustrated on different scales in (d).   
\label{fig:DFTvsEXP}}
\end{figure*}

We use machine learning to explore the MAE of the $L1_0$ phase, resulting in a smooth profile in the $\mathbb{R}^{5}$ compositional space that is consistent with the experimental trend.  
Similar to the previous cases, we first modulate the compositions near the equiatomic case for the transition metals, while keeping Pt fixed at $50\%$. 
For all explored cases, DFT suggests that the $[001]$ direction is an easier axis compared to $[100]$, and the MAE values are shown in Fig. \ref{fig:DFTvsEXP}(a).  
Although this result suggests significant correlation between the Fe concentration and the MAE, it is clear that all transition-metal elements have strong influences. 
To fully capture and visualize the MAE profile, we use machine learning and interpolation in the compositional space. 
This is implemented using multidimensional scaling (MDS), an unsupervised learning method typically used to cluster high-dimensional points \cite{kruskal_nonmetric_1964}.
MDS searches for an optimal curved 2-dimensional surface in a high (5 in our case) dimensional parameter space, such that the points can uniformly scale their Euclidean distance to their neighbors after projected onto the surface. 
This surface is numerically described by a discrete manifold, which can be illustrated on a flat 2D plane after a straightforward transformation. 

The flattened MDS map of the data points in Fig.\ref{fig:DFTvsEXP}(a) is illustrated in Fig.\ref{fig:DFTvsEXP}(b), with the projected locations denoted by the crosses. 
The MAE values obtained by DFT are illustrated with the white-salmon color scale, where the empty space between points are filled using Hermite-Zero-Derivative Start (HZDS) interpolation. 
The DFT results are compared with experiments using five different $L1_0$ thin films near the equiatomic composition. 
These samples were fabricated and characterized using the same approach described in our previous work\cite{beeson_single_2023}, where the compositions are estimated by Energy-Dispersive X-Ray Spectroscopy (EDS) and the MAE values estimated by magnetometry measurements. 
The specific compositions of these five samples are listed in Table \ref{tab:ElementalCompositions}. 
\begin{table}
\centering
\caption{Elemental compositions of the samples. \label{tab:ElementalCompositions}}
\begin{tabular}{cc} 
\hline 
Samples & Compositions \\
\hline 
\hline 
1 & $\textrm{Fe}_{0.07}\textrm{Co}_{0.12}\textrm{Ni}_{0.11}\textrm{Mn}_{0.11}\textrm{Cu}_{0.12}\textrm{Pt}_{0.47}$ \\
2 & $\textrm{Fe}_{0.10}\textrm{Co}_{0.08}\textrm{Ni}_{0.20}\textrm{Mn}_{0.07}\textrm{Cu}_{0.09}\textrm{Pt}_{0.46}$ \\
3 & $\textrm{Fe}_{0.11}\textrm{Co}_{0.09}\textrm{Ni}_{0.10}\textrm{Mn}_{0.09}\textrm{Cu}_{0.11}\textrm{Pt}_{0.50}$ \\
4 & $\textrm{Fe}_{0.13}\textrm{Co}_{0.11}\textrm{Ni}_{0.07}\textrm{Mn}_{0.09}\textrm{Cu}_{0.11}\textrm{Pt}_{0.49}$ \\
5 & $\textrm{Fe}_{0.19}\textrm{Co}_{0.10}\textrm{Ni}_{0.10}\textrm{Mn}_{0.07}\textrm{Cu}_{0.08}\textrm{Pt}_{0.46}$ \\
\hline 
\end{tabular}
\end{table}
These compositions are projected to the same MDS manifold, which are illustrated by the solid circles in Fig. \ref{fig:DFTvsEXP}(b) where the white-blue scale represents the normalized experimental values. 
It can be seen that the general trend of the experimental MAE is highly correlated to the interpolated MDS profile.  
We also performed DFT calculations using the experimental compositions of the five samples, and the results of $M_S$ and MAE are illustrated in Figs. \ref{fig:DFTvsEXP}(c-d). 
In both cases, DFT overestimates the values by at least several factors, which is consistent with the trend seen in the case of pure FePt (Table \ref{tab:Comparison}).  
This may be induced by the assumption of perfect crystallization, whereas the real samples have complicated grain landscapes and lattice distortions that are difficult to capture by the DFT setup.
The overestimation is also likely induced by the lost information of the mean-field approach in CPA, where consecutive scattering is ignored.
Although quantitative agreement between theory and experiment is difficult, the trends of $M_S$ and MAE are both consistent with the experimental ones. 
Note that the experimental compositions are not used in the interpolation shown in Fig. \ref{fig:DFTvsEXP} (b), suggesting that the smooth profile of the MAE can be captured by the sparse DFT points sampled in the compositional space. 
The results shown above suggest that sparse sampling in the compositional space of HEAs using primitive-cell DFT calculations can capture the trend of magnetic properties including $M_S$, MAE and $T_C$, even after losing the information of atomic occupation details within CPA. 
Since CPA assumes translational symmetry, the Green's functions are naturally diagonal in $\mathbf{k}$, allowing for the fast evaluation of the electron density by working only within the first Brillouin zone. 
Although the experiments do not quantitatively agree with DFT, the profile in the high-dimensional compositional space is smooth, which can be captured by a reasonable number of discrete points sparsely sampled in the compositional space. 
This invites further research to establish large-scale databases within the mean-field approximation of CPA to establish predictive machine-learning models. 
This also introduces an intriguing opportunity to achieve the fast design of future high-entropy magnets for many scenarios of application. 
\emph{Acknowledgments:} This paper is based upon work supported by the National Science Foundation (US) under Grant No. ECCS-2151809. This work used Bridges-2 at Pittsburgh Supercomputing Center through allocation PHY230018 from the Advanced Cyberinfrastructure Coordination
Ecosystem: Services \& Support (ACCESS) program, which is supported by National Science Foundation (US) grants \#2138259, \#2138286, \#2138307, \#2137603, and \#2138296. The acquisition of the MPMS3 system used in this study was supported by the NSF-MRI program (DMR-1828420). TS and SNK gratefully acknowledge funding from US Department of Energy (DOE), under the award DE-SC0006420. JJ acknowledges support under the CCP9 project "Computational Electronic Structure of Condensed Matter"(part of Computational Science Centre of Research Communities(CoSeC)).

\bibliographystyle{apsRevNoPublisherFullAuthorFullTitle.bst}
\bibliography{HEA.bib,HEA2.bib}

\end{document}